\documentclass{llncs}
\usepackage[symbol]{footmisc}
\usepackage{graphicx}
\usepackage{booktabs}
\usepackage{wrapfig}
\usepackage[table]{xcolor}
\usepackage{nicefrac}
\usepackage{algorithm2e}
\usepackage{hyperref}       
\usepackage{amsfonts}       
\usepackage{amsmath}
\usepackage{microtype}

\begin{document}
\title{Classifying Malware Using Function Representations in a Static Call Graph}

\author{Thomas Dalton \and 
        Mauritius Schmidtler \and 
        Alireza Hadj Khodabakhshi}

\institute{Webroot \footnote{Webroot is an OpenText company.} \\ \email{\{tdalton,mschmidtler,ahadjkhodaba\}@opentext.com}}

\maketitle
\begin{abstract}
    We propose a deep learning approach for identifying malware families using the function call graphs of x86 assembly instructions.
    Though prior work on static call graph analysis exists, very little involves the application of modern, principled feature learning techniques to the problem.
    In this paper, we introduce a system utilizing an executable's function call graph where function representations are obtained by way of a recurrent neural network (RNN) autoencoder which maps sequences of x86 instructions into dense, latent vectors. 
    These function embeddings are then modeled as vertices in a graph with edges indicating call dependencies.
    Capturing rich, node-level representations as well as global, topological properties of an executable file greatly improves malware family detection rates and contributes to a more principled approach to the problem in a way that deliberately avoids tedious feature engineering and domain expertise.
    We test our approach by performing several experiments on a Microsoft malware classification data set and achieve excellent separation between malware families with a classification accuracy of 99.41\%.

\end{abstract}

\section{Introduction}
Malware is often classified into families based on certain shared characteristics between samples.
It is often very useful to distinguish between malware families in order to detect trends in malware infections over time and to attribute authorship.
Traditionally, classifying malware has required teams of threat researchers to perform advanced reverse engineering techniques in order to identify various unique characteristics that define a family.
However, cyber threats have exploded in recent years making it difficult for threat researchers to keep up.
Malware in particular continues to grow in sophistication with new strains released daily.
The practice of malware polymorphism renders traditional automated signature-based approaches ineffective for identifying novel instances of malware.

\begin{figure}
    \centering
    \includegraphics[width=.5\linewidth]{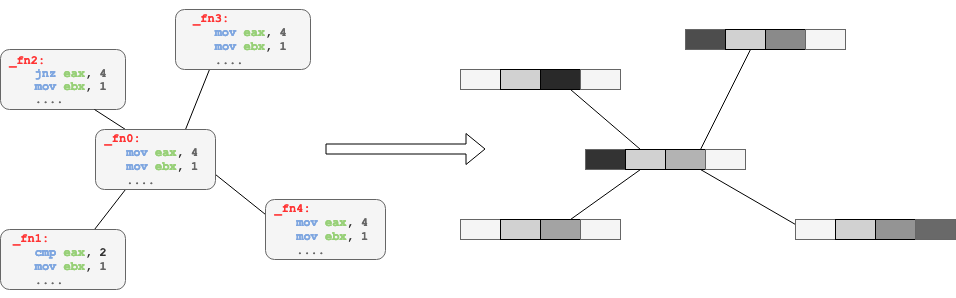}
    \caption{Variable-length sequences of x86 instructions found in functions are embedded into fixed-length vectors using a GRU-based sequence-to-sequence autoencoder.}
\end{figure}

In this work, we propose a new approach to malware classification that is inspired by reverse engineering techniques yet requires no domain-specific feature engineering and is invariant to polymorphism.
Specifically, we devise a function call-graph framework in which the function representations are learned.
By framing the problem through the lens of representation learning, we are able to greatly improve automatic classification while also contributing to human insight which is helpful to determine authorship and intent.
By incorporating rich, node-level representations as well as global, structural properties of an executable's call graph, we are able to classify malware families with very high accuracy.
Our approach consists of several composite models aimed at learning robust function representations that when employed together form the full classification system.

\section{Related Work}
Call graphs are commonly used by malware analysts and reverse engineers to manually analyze and inspect executable files \cite{p158-murphy}.
Indeed, many real-world data is naturally represented using graphs.
Graphs have been successfully utilized in analyzing data from a wide variety of domains including social network link prediction \cite{liben-nowell2007link-pred}, protein-protein interactions \cite{airola2008gkp}, and communication networks \cite{mesbahi2010graph}. 
Due to their expressive ability, there is growing interest in applying machine learning techniques directly to graph-represented data to bypass tedious feature engineering.
Graph kernels have been proposed to allow for kernel-based methods (such as support vector machines) to be applied directly to graph classification problems.
Kernels based on the Weisfeiler-Lehman test of graph isomorphism have grown in popularity in recent years \cite{shervashidze2011weisfeiler} owing to their relative simplicity and strong discriminative ability.

In prior works, call graphs have been used to automatically classify malware but typically these works employ relatively simple graph similarity measures such as graph edit distance or rely on heavy feature engineering involving summary statistics to describe functions in the graph \cite{hassen2017scalable,searles2017parallelization,dullien2005,Dam2017MalwareDB}. 
We build on this call graph approach by incorporating certain representation learning techniques such as autoencoding and clustering \cite{Goodfellow:2016:DL:3086952} to obtain an improved function representation.
By extending the well-established call graph strategies with a principled representation learning approach, we forego the tedious and heuristic feature engineering steps of prior work, giving us much better graph representations.

\begin{figure}
    \centering
    \includegraphics[width=0.3\linewidth]{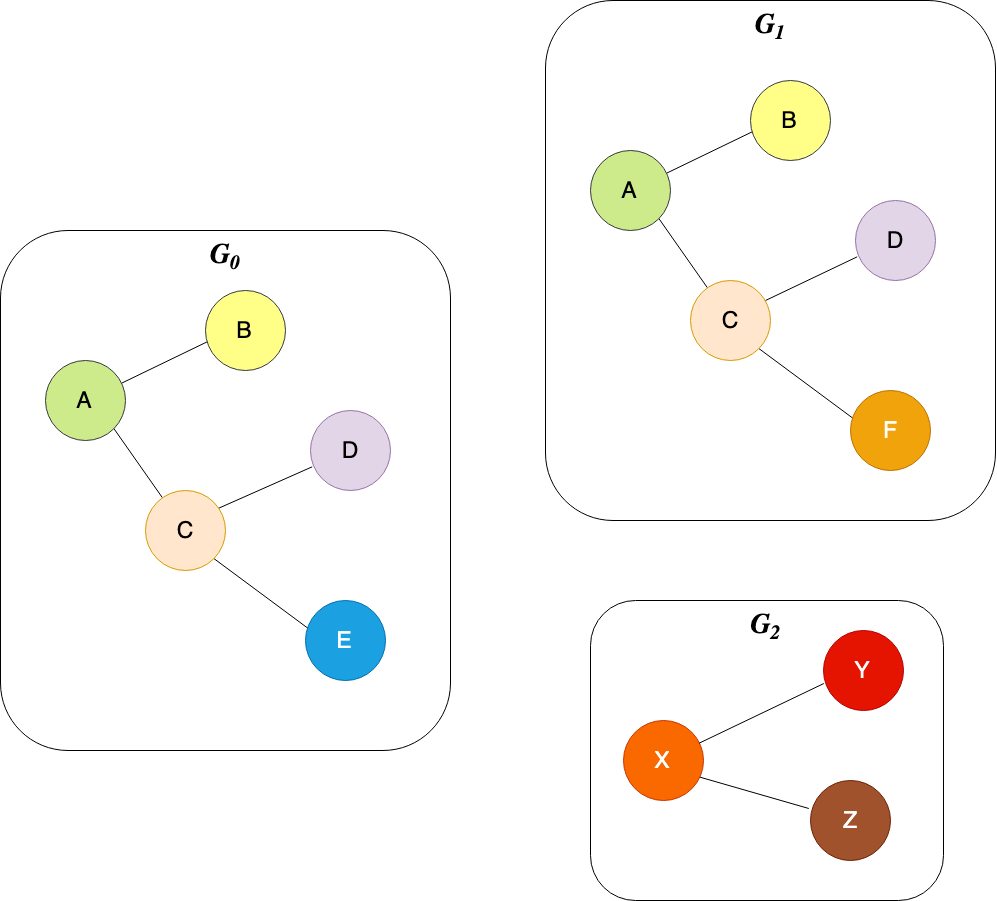}
    \caption{$G_0$ and $G_1$ are considered similar since they both share a similar graph topology and near-identical vertex label sets.
    On the other hand, $G_2$ is considered very different from both $G_0$ and $G_1$ since it neither shares a graph topology nor a vertex set with the other graphs.
    The notion of graph ``similarity'' is clarified in Section \ref{graph-section}}
    \label{graph-intuition}
\end{figure}

\section{Call Graph Framework}
A call graph describes the logical control flow of an executable file where functions (or subroutines) are expressed as vertices and edges represent a dependency or call relationship.
Call graphs have proven to be extremely useful to security researchers for the analysis and classification of malicious software.
\begin{wrapfigure}{l}{0.5\textwidth}
    \centering
    \begin{tabular}{ll}
        \toprule
        Actual & Predicted \\
        \midrule
        \cellcolor{gray!5} \texttt{mov edi, edi} & \cellcolor{green!30} \texttt{mov edi, edi} \\
        \cellcolor{gray!5} \texttt{push ebp} & \cellcolor{green!30} \texttt{push ebp} \\
        \cellcolor{gray!5} \texttt{mov ebp, esp} & \cellcolor{green!30} \texttt{mov ebp, esp} \\
        \cellcolor{gray!5} \texttt{sub esp, 48} & \cellcolor{red!5} \texttt{sub esp, \textcolor{red}{24}} \\
        \rowcolor{gray!5} \texttt{lea ecx, 0x} & \cellcolor{green!30} \texttt{lea ecx, 0x} \\
        \rowcolor{gray!5} \texttt{push eax} & \cellcolor{red!30} \texttt{<unknown>} \\
        \rowcolor{gray!5} \texttt{push ecx} & \cellcolor{green!30} \texttt{push ecx} \\
        \rowcolor{gray!5} \texttt{call <addr>} & \cellcolor{green!30} \texttt{call <addr>} \\
        \rowcolor{gray!5} \texttt{pop ecx} & \cellcolor{green!30} \texttt{pop ecx} \\
        \rowcolor{gray!5} \texttt{\dots} & \cellcolor{green!10} \texttt{\dots} \\
        \rowcolor{gray!5} \texttt{\dots} & \cellcolor{green!10} \texttt{\dots} \\
        \bottomrule \\
    \end{tabular}
    \caption{\label{fig:decoder-example}
            An example sequence decoding given a latent embedding.
            The decoder is able to re-create the original sequence with high accuracy, indicating that the latent embedding has captured sufficient information.}
\end{wrapfigure}
The main intuition behind the call graph approach to malware classification is that files sharing similar call graphs are likely to have been generated from the same family.
By understanding the logical flow of the executable, we can gain significant insight into the intent of the malware.
It is important, therefore, to represent the graph such that we capture rich vertex-level information as well as global, topological properties of the graph.
The intuition is illustrated in Figure \ref{graph-intuition}.

\subsection{Overview}
We break the malware classification task down into three primary subtasks which we summarise here.
In order to obtain a good whole-graph representation of the executable, it is important to first obtain high quality embeddings for the functions contained within the file.
For this, we use a sequence-to-sequence autoencoder which captures the sequential nature of the x86 code instructions into a low-dimensional, latent representation of the function.

This function embedding helps to make our model more robust to polymorphic techniques since a perturbation in the x86 instruction space results in a proportional perturbation in the embedding space.
It can also be useful for identifying the specific functions that make the file malicious.
This function embedding approach is one of the key differentiators between our approach and prior call graph approaches to malware classification.

\begin{wrapfigure}{r}{0.25\textwidth}
    \includegraphics[width=0.2\textwidth]{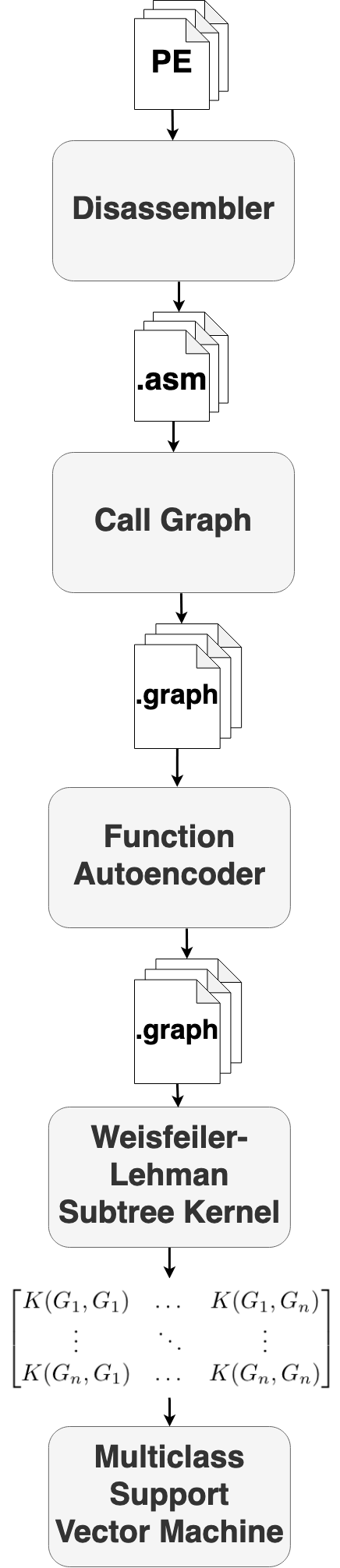}
    \label{seq2seq-fig}
\end{wrapfigure}

Having obtained function representions, we cluster the embeddings to obtain discrete labels for the functions and re-label the graph vertices according to their respective cluster IDs.
Finally, the whole-graph representation is obtained using a graph kernel inspired by the Weisfeiler-Lehman test of graph isomorphism.
The message-passing property of the Weisfeiler-Lehman framework allow us to efficiently capture the global structure of the graph.

The executable binary files are disassembled into plain text \texttt{.asm} files using IDA \cite{hexrays2011ida}, a popular disassembler widely used by security researchers.
Due to the tendency of code sections to contain very long sequences (sometimes upwards of hundreds of thousands of instructions), we break up the sequences into functions or subroutines which provide natural delimiters much like sentences and paragraphs are in a document.
These shorter length sequences enable us to use recurrent neural units such as long short-term memory (LSTM) or gated recurrent units (GRU) where the training samples are individual functions with sequences ranging from very short (fewer than 5 instructions) to as long as a few hundred instructions. 
By following the \texttt{call} instructions in the code, we can construct the file's call graph, $G = (\mathcal{V}, \mathcal{E})$ where vertices $v\in\mathcal{V}$ represent functions and edges $e=(v,v')\in\mathcal{V} \times \mathcal{V}$ represent a call dependency. 

\begin{figure}
    \centering
    \includegraphics[width=0.5\linewidth]{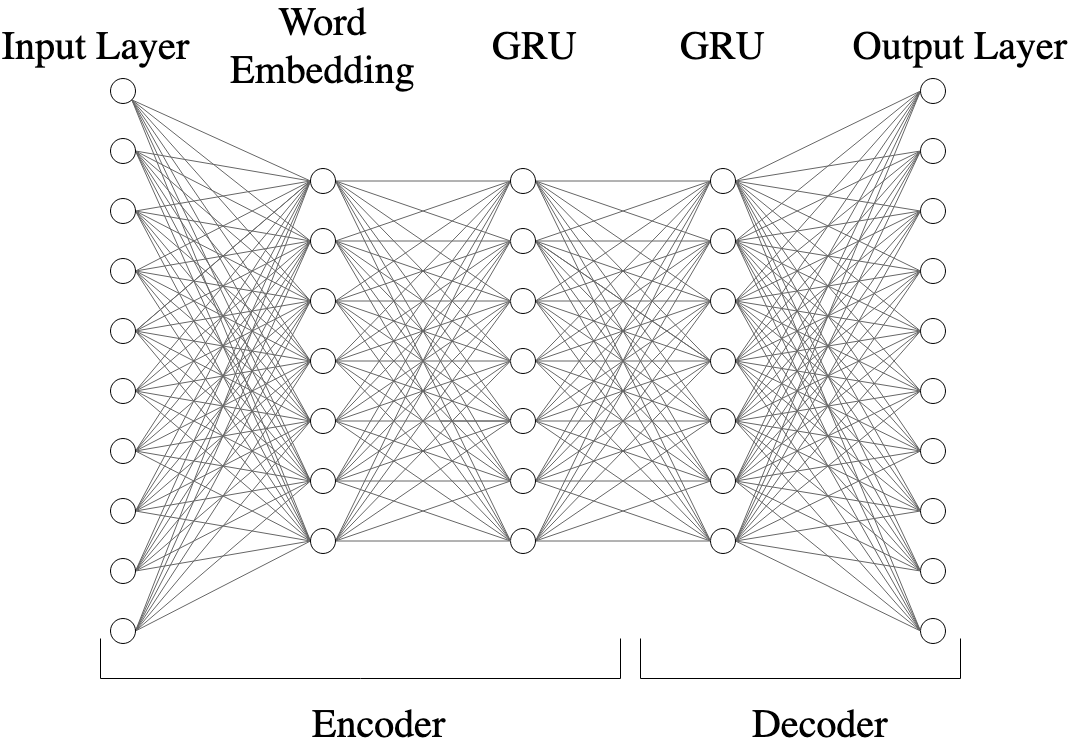}
    \caption{The sequence-to-sequence autoencoder architecture. The left side of the network encodes the sequence into a fixed-length latent representation.
    The right side of the network decodes the latent representation back into the original sequence.}
    \label{seq2seq-fig}
\end{figure}

The model considers only portions of the executable containing valid x86 code instructions.
Traditionally, these code sections are identified by their section header name (e.g. \texttt{.text}, \texttt{.code}, etc.) but malware often obfuscates intent by using a packer which may result in non-standard section names such as \texttt{.brick} or \texttt{iuagwws}. 
In addition to all code found in the standard code sections, our approach also considers such nonstandard sections containing valid x86 code to construct the call graph.

We make a distinction between two kinds of vertices: internal functions and external functions.
Internal functions are those that are present in the executable and subsequently can be disassembled directly.
External functions are those which are imported from external libraries and thus the code is not readily available for disassembly.
Our graph therefore consists of edges between both internal and external functions.
It is worth noting that executable files also contain sections that do not typically contain code such as \texttt{.data} or \texttt{.reloc}.
While these sections provide additional data that is often quite useful for malware classification, we ignore any non-code data for the call graph construction task.
It is possible to attribute the graph with the information contained in such non-code sections but that is beyond the scope of this work as our principle concern is that of malware classification using x86 code representations.

A sequence-to-sequence \cite{sutskever2014sequence} GRU-based \cite{cho2014gru} autoencoder architecture was chosen for the task of embedding variable-length sequences of x86 code instructions into fixed-length, continuous vectors. 
The function embedding model is comprised of an encoder and a decoder with the encoder being responsible for compressing sequences into low-dimensional, latent representations. 
A decoder is used to decompress the fixed-length vector back into the original variable-length sequence.
Because the autoencoder model must recreate the original sequence from its bottleneck representation, the model learns an efficient, latent representation of the original sequence.
After the autoencoder model is trained, the decoder is discarded and only the encoder portion is used to encode new sequences.
Sequence-to-sequence architectures have been used successfully in language modeling tasks such as machine translation.
Often, the goal is to translate a sequence of words from one language, such as English, into another language, such as French.
\begin{figure}
    \centering
    \includegraphics[width=0.5\linewidth]{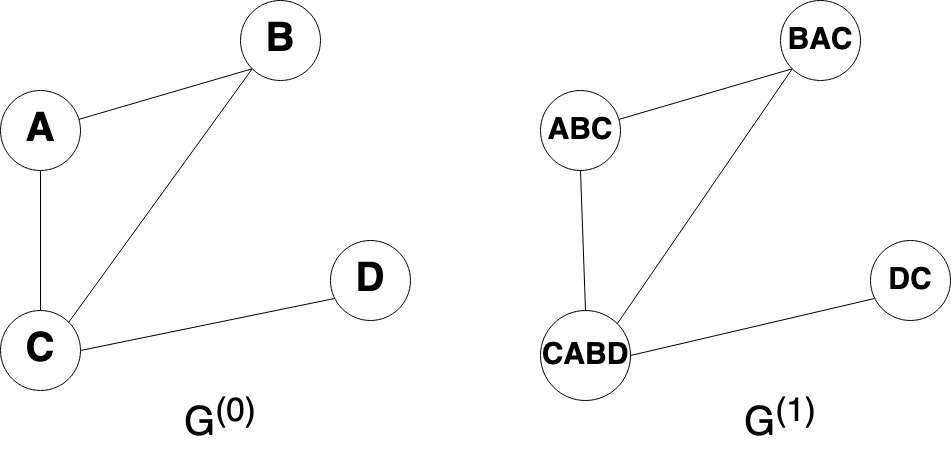}
    \caption{For each iteration of the Weisfeiler-Lehman algorithm, a new label is generated for each vertex.
    The new label is derived by concatenating the labels of the adjacent vertices.}
    \label{wl-illustration}
\end{figure}
The input sequence is encoded into a bottleneck layer which captures a latent representation of the sequence irrespective of language.
In machine translation tasks, the input and output sequences are usually composed of words drawn from disjoint vocabularies.
However, our sequence-to-sequence task involves reconstructing the original input sequence from the bottleneck representation, so the same vocabulary is used for both the input and the output sequences.
By reconstructing the original input sequence from the latent representation, the sequence-to-sequence network becomes an autoencoder.

\subsection{Function Representations}

Given a GRU with the following definitions,

\begin{align*}
    z_t &= \sigma_g(W_z x_t + U_z h_{t-1} + b_z) \\
    r_t &= \sigma_g(W_r x_t + U_r h_{t-1} + b_r) \\
    h_t &= (1 - z_t) \circ h_{t-1} + z_t \circ \sigma_h(W_h x_t + U_h(r_t \circ h_{t-1}) + b_h)
\end{align*}

the hidden representation of a sequence with length $T$ is taken to be $h_T$. 

During training, when the encoder receives the last token in the sequence $x_T=$ \texttt{<end>}, the decoder is initialized with $h_{T}^{(enc)}$, thus transferring the compressed sequence information to the decoder.
In addition to the final hidden state of the encoder, the decoder is also supplied with the original input sequence with a one-step delay.
That is, at time step $t$, the decoder receives the true input $x_{t-1}$ for $t>1$ where $x_{0}=$ \texttt{<start>}.
This technique of supplying the original sequence with a delay into the decoder is known as teacher forcing.

The decoder, therefore, is trained to predict the next token of the sequence given the hidden state of the encoder and the previous time step.
This helps to greatly speed up the training of the autoencoder.
An example decoding is illustrated in Figure \ref{fig:decoder-example}.
After training is completed, the decoder portion of the model is thrown away and only the encoder is used to obtain latent representations for function sequences.

It is common for authors of malware to obfuscate the intent of a file by adding junk instructions such as no-op instructions. 
Because the function is represented in a latent space, it is relatively immune to such common obfuscation tactics which can often thwart signature-based or count-based solutions. 

In our model, input sequences are tokenized as the full x86 assembly instruction after replacing relative memory address locations with an \texttt{<addr>} string.
The vocabulary is thus composed of the top 20,000 most common x86 instructions found in the corpus plus the \texttt{<start>} and \texttt{<end>} tokens.

\RestyleAlgo{ruled}
\SetKwComment{Comment}{$\triangleright$\ }{}
\begin{wrapfigure}{L}{0.5\textwidth}
\begin{algorithm}[H]
    \DontPrintSemicolon
    \caption{Weisfeiler-Lehman Subtree Graph Kernel}
    \label{wl-alg}
    \KwIn{$G$, $G'$, $h$}
    \KwOut{$k^{(h)}(G, G')$}
    \For{$i\leftarrow 1$ \KwTo $h$} {
        \For{$v \in \mathcal{V}$} {
            $N_v=neighbors(v)$ \\ 
            $S_v=sort(N_v)$ \\ 
            $v, \sigma_v=hash(S_v)$ 
        }
        \For{$v' \in \mathcal{V'}$} {
            $N_{v'}=neighbors(v')$ \\ 
            $S_{v'}=sort(N_{v'})$ \\ 
            $v', \sigma_{v'}=hash(S_{v'})$ 
        }
    }
    \;
    $\phi(G) = \left[c(G, \sigma_0), \dots, c(G, \sigma_s)\right]$\;
    $\phi(G') = \left[c(G', \sigma_0), \dots, c(G', \sigma_s)\right]$\;
    \;
    \Return{$\langle\phi(G), \phi(G')\rangle$}
\end{algorithm}
\end{wrapfigure}

The sequence-to-sequence autoencoder network is constructed as in Figure \ref{seq2seq-fig} with three hidden layers -- two in the encoder and one in the decoder.
The first hidden layer in the encoder is a fully connected layer for learning x86 instruction embedding vectors of length 64.
The weights in this layer can be pre-trained using a Word2vec \cite{mikolov2013efficient}, GloVe \cite{pennington2014glove}, or some other similar unsupervised word embedding technique.
However in our experiments we observed no ill effects from initializing the weights randomly and learning the instruction embedding directly as part of the autoencoder training. 

\subsection{Function Clustering}
Having obtained an x86 instruction sequence encoder, we embed all internal functions found in the function call graph and attribute the vertices with their respective embeddings.
External vertices representing imported functions or APIs remain non-attributed since we are unable to obtain function embeddings for these functions.
They therefore retain their discrete external labels.
The graphs then are composed of two types of vertices: internal functions with continuous attributes but arbitrary discrete labels and external functions with discrete labels but no continuous attributes.
In order to carry out the graph classification task with established Weisfeiler-Lehman kernels, we must first obtain learned discrete labels for internal functions.
We achieve this by clustering their continuous embeddings and labeling clusters according to their cluster identifier.
We can then label the internal functions in the graph with their discrete cluster identifier.
For the purposes of graph classification, our labels can be non-descriptive (e.g. $C_1, C_2,...$) but it may be beneficial in future work to consider assigning descriptive cluster labels to aid humans in manual threat research analysis.

Even modestly-sized files can contain tens of thousands of individual functions so it is important to be able to scale our clustering task for many function samples.
Thus, we use the popular mini-batch K-means clustering algorithm \cite{sculley2010kmeans} as it is easily able to scale to many millions of samples.
Using the so-called ``elbow method'', we found that $k=7000$ was a reasonable choice for the number of clusters for our data sets.
Although several hierarchical- and density-based clustering algorithms have been found to yield superior results over K-means, these algorithms are generally unable to scale to many samples. 

\subsection{Graph Classification} \label{graph-section}
A graph kernel is a function that computes an inner product between graphs and can be thought of as a way to measure graph similarity.
Graph kernels are widely studied since they allow kernel-based machine learning algorithms such as SVMs to be applied directly to graph-structured data.
Most graph kernels are based on the Weisfeiler-Lehman test of graph isomorphism.
Indeed, we employ the popular Weisfeiler-Lehman subtree kernel algorithm to obtain the whole-graph feature vector.
Once a feature vector for the graph is obtained, we can compute a pairwise kernel matrix and train an SVM for malware classification.

Let the Weisfeiler-Lehman kernel with base kernel $k$ be defined as 

$$K^{(h)}_{WL}(G,G') = \sum_{i=0}^{h}{\alpha_i k(G_i, G'_i)}$$

where $\{G_0, G_1, ..., G_h\}$ and $\{G'_0, G'_1, ..., G'_h\}$ are sequences of graphs that the Weisfeiler-Lehman algorithm generates from $G$ and $G'$  respectively after $h$ iterations. 
For each iteration of the Weisfeiler-Lehman algorithm, each vertex obtains a new label by concatenating the labels of the adjacent vertices. 
This process is illustrated for one iteration of the algorithm in Figure \ref{wl-illustration}.
The Weisfeiler-Lehman kernel, then, is simply the weighted sum of the base kernel function applied to the graphs generated by the Weisfeiler-Lehman algorithm.
Let the subtree base kernel $k$ be defined as the inner product between $\phi(G)$ and $\phi(G')$

$$k(G, G')=\langle\phi(G), \phi(G')\rangle$$

where $\phi(Q) = \left[ c(Q,\sigma_{0}), c(Q,\sigma_{1}), ..., c(Q,\sigma_{s})\right]$ and $c(Q, \sigma_i)$ is the count of vertex label $\sigma_{i}\in\Sigma^{(h)}$ occurring in the graph $Q$.
The set of vertex labels obtained after $h$ iterations of the Weisfeiler-Lehman algorithm is denoted as $\Sigma^{(h)}$.
To compute the kernel matrix, we compute the pairwise kernels for all graphs as 
$$
K^{(h)}=
    \begin{bmatrix}
        k^{(h)}(G_1, G_1) & \dots & k^{(h)}(G_1, G_N) \\
        k^{(h)}(G_2, G_1) & \dots & k^{(h)}(G_2, G_N) \\
        \vdots & \ddots & \vdots \\
        k^{(h)}(G_N, G_1) & \dots & k^{(h)}(G_N, G_N)
    \end{bmatrix}
$$

Shervashidze, et al (2011) have shown that for $N$ graphs with $n$ vertices and $m$ edges, the Weisfeiler-Lehman subtree kernel of height $h$ can be computed in $O(Nhm + N^2hn)$ time. 
This kernel matrix can be supplied directly to a one-versus-all support vector machine with Platt scaling in order to obtain the class probabilities \cite{platt1999proba}.

\section{Experiments}

\subsection{Setup}
We performed three separate end-to-end experiments of our malware classifier system which is composed of the individually trained components below.
Each component feeds into the next to form the final multiclass classifier system.

\begin{itemize}
    \item Sequence-to-sequence autoencoder model
    \item K-means clustering model
    \item Weisfeiler-Lehman subtree kernel model
\end{itemize}

\begin{table}
    \caption{Malware descriptions}
    \centering
    \begin{tabular}{lcl}
        \toprule
        Family Name & Samples & Type \\
        \midrule
        Ramnit & 1263 & Worm \\ 
        Lollipop & 2306 & Adware \\
        Kelihos\_ver3 & 2931 & Backdoor \\
        Vundo & 344 & Trojan \\
        Simda & 33 & Backdoor \\ 
        Tracur & 663 & TrojanDownloader \\ 
        Kelihos\_ver1 & 382 & Backdoor \\
        Obfuscator.ACY & 1158 & Obfuscated Malware \\ 
        Gatak & 954 & Backdoor \\
        \bottomrule
    \end{tabular}
\end{table}

Ten percent of the original data set was withheld from training altogether, not seen by any of the individual sub-models and was used only for testing the composite classifier models. 
Ten percent of the training set was used for validation.
Deep learning library Keras \cite{chollet2015keras} was used to construct and train the sequence-to-sequence network.
The network was trained on an NVIDIA Tesla K80 GPU.
We used the GraKeL \cite{siglidis2018grakel} implementation of the Weisfeiler-Lehman algorithm with three iterations to obtain the Weisfeiler-Lehman graphs and the kernel matrix.
The scikit-learn \cite{scikit-learn} implementation for support vector machines based on LIBSVM \cite{libsvm} was used for the graph classification task with hyperparameters $C$ and $\gamma$ being obtained through grid search.

\begin{table*}
    \caption{Results summary}
    \centering
    \label{summary}
    \begin{tabular}{l c c c c c c c c }
        \toprule
        Family Name & Precision & Recall & F1-Score & TP & FP & FN & TN & Support \\
        \midrule
        Ramnit & 0.970 & 0.995 & 0.982 & 382 & 12 & 2 & 2682 & 384 \\ 
        Lollipop & 0.997 & 1.000 & 0.998 & 708 & 2 & 0 & 2358 & 708 \\ 
        Kelihos\_ver3 & 1.000 & 1.000 & 1.000 & 861 & 0 & 0 & 2205 & 861 \\ 
        Vundo & 0.989 & 1.000 & 0.994 & 93 & 1 & 0 & 2973 & 93 \\ 
        Simda & 1.000 & 1.000 & 1.000 & 15 & 0 & 0 & 3051 & 15 \\ 
        Tracur & 0.984 & 1.000 & 0.992 & 180 & 3 & 0 & 2886 & 180 \\
        Kelihos\_ver1 & 1.000 & 0.967 & 0.983 & 87 & 0 & 3 & 2979 & 90 \\ 
        Obfuscator.ACY & 1.000 & 0.974 & 0.987 & 368 & 0 & 10 & 2698 & 378 \\
        Gatak & 1.000 & 0.992 & 0.996 & 354 & 0 & 3 & 2712 & 357 \\ 
        \bottomrule
    \end{tabular}
\end{table*}

\subsection{Data Set}

We use the Microsoft Malware Classification data set \cite{2018arXiv180210135R} to evaluate our approach. 
The data set consists of samples from nine different malware families. 
Each sample in the data set is composed of a pre-disassembled \texttt{.asm} file generated by IDA and a sanitized hexadecimal representation of the original executable. 
Our approach makes use of only the \texttt{.asm} and furthermore only takes advantage of the parsable code sections.
Having obtained only the publicly available training set of 10,867 samples, we were able to extract function call graphs for 10,152 samples due, in part, to certain samples being packed or otherwise obfuscated.

\subsection{Results}
We achieved a prediction accuracy of 99.41\% in the malware family classification task.
Our approach outperforms other malware classifiers that involve extensive feature engineering or extract significantly more data from the executable such as non-code data \cite{cics17,hassen2017scalable,searles2017parallelization}.
Since we only use the code sections of the executable, we expect that incorporating additional data such as the \texttt{.rsrc} and \texttt{.idata} sections would help to further improve classification results.
Table \ref{summary} summarizes the results across the three experiments.

\section{Conclusion}
In this work we applied several machine learning techniques to the problem of malware detection and achieved over 99\% accuracy in the malware classification task using a composite model.
A sequence-to-sequence autoencoder was used to obtain dense, latent representations of x86 code which helped our model account for anti-malware evasion practices.
We then clustered the function representations of the functions and obtained discrete function labels.
Using the discrete labels, we constructed a function call graph where vertices represent functions and are labeled according to their cluster IDs.
The Weisfeiler-Lehman graph kernel framework was used to obtain the Weisfeiler-Lehman graphs and to construct a kernel matrix which allowed us to ultimately perform the graph classification task using a support vector machine.

\section{Acknowledgments}
Special thanks to our colleague, Andrew Sandoval, for his valuable contributions and feedback throughout this work.

\begin{figure}[h]
    \centering
    \label{confusion-matrix}
    \caption{Normalized confusion matrix}
    \includegraphics[width=0.8\linewidth]{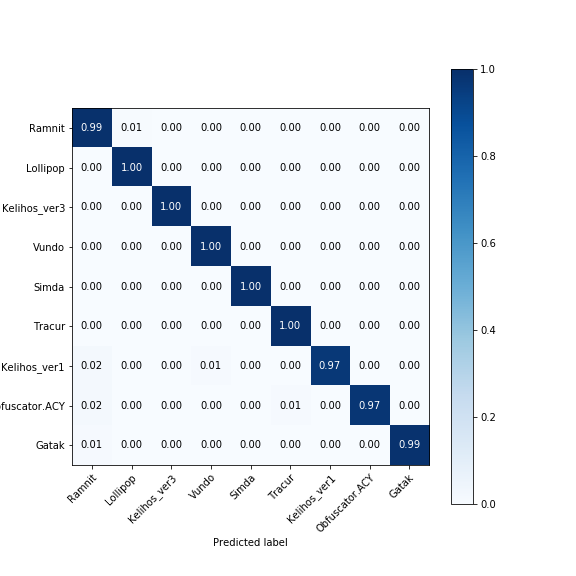}
\end{figure}

\bibliographystyle{splncs04}
\bibliography{biblio}

\begin{thebibliography}{10}
\providecommand{\url}[1]{\texttt{#1}}
\providecommand{\urlprefix}{URL }
\providecommand{\doi}[1]{https://doi.org/#1}

\bibitem{airola2008gkp}
Airola, A., Pyysalo, S., Bj\"{o}rne, J., Pahikkala, T., Ginter, F., Salakoski,
  T.: A graph kernel for protein-protein interaction extraction. In:
  Proceedings of the Workshop on Current Trends in Biomedical Natural Language
  Processing. pp.~1--9. BioNLP '08, Association for Computational Linguistics,
  Stroudsburg, PA, USA (2008),
  \url{http://dl.acm.org/citation.cfm?id=1572306.1572308}

\bibitem{libsvm}
Chang, C.C., Lin, C.J.: {LIBSVM}: A library for support vector machines. ACM
  Transactions on Intelligent Systems and Technology  \textbf{2},  27:1--27:27
  (2011), software available at \url{http://www.csie.ntu.edu.tw/~cjlin/libsvm}

\bibitem{cho2014gru}
{Cho}, K., {van Merrienboer}, B., {Gulcehre}, C., {Bahdanau}, D., {Bougares},
  F., {Schwenk}, H., {Bengio}, Y.: {Learning Phrase Representations using RNN
  Encoder-Decoder for Statistical Machine Translation}. arXiv e-prints
  arXiv:1406.1078 (Jun 2014)

\bibitem{chollet2015keras}
Chollet, F., et~al.: Keras. \url{https://keras.io} (2015)

\bibitem{Dam2017MalwareDB}
Dam, K.H.T., Touili, T.: Malware detection based on graph classification. In:
  ICISSP (2017)

\bibitem{dullien2005}
Dullien, T.: Graph-based comparison of executable objects (2005)

\bibitem{Goodfellow:2016:DL:3086952}
Goodfellow, I., Bengio, Y., Courville, A.: Deep Learning. The MIT Press (2016)

\bibitem{cics17}
{Hassen}, M., {Carvalho}, M.M., {Chan}, P.K.: Malware classification using
  static analysis based features. In: 2017 IEEE Symposium Series on
  Computational Intelligence (SSCI). pp.~1--7 (Nov 2017).
  \doi{10.1109/SSCI.2017.8285426}

\bibitem{hassen2017scalable}
Hassen, M., Chan, P.: Scalable function call graph-based malware
  classification. In: Proceedings of the Seventh ACM on Conference on Data and
  Application Security and Privacy, CODASPY ’17. pp. 239--248 (Mar 2017).
  \doi{10.1145/3029806.3029824}

\bibitem{hexrays2011ida}
Hex-Rays: The ida pro disassembler and debugger.
  \url{https://www.hex-rays.com/products/ida/} (Jan 2011)

\bibitem{liben-nowell2007link-pred}
Liben-nowell, D., Kleinberg, J.: The link prediction problem for social
  networks. Journal of the American Society for Information Science and
  Technology  \textbf{58} (Jan 2003). \doi{10.1002/asi.20591}

\bibitem{mesbahi2010graph}
Mesbahi, M., Egerstedt, M.: Graph theoretic methods in multiagent networks.
  Princeton University Press (2010)

\bibitem{mikolov2013efficient}
Mikolov, T., Chen, K., Corrado, G., Dean, J.: Efficient estimation of word
  representations in vector space. arXiv preprint arXiv:1301.3781  (2013)

\bibitem{p158-murphy}
Murphy, G.C., Notkin, D., Griswold, W.G., Lan, E.S.: An empirical study of
  static call graph extractors. ACM Trans. Softw. Eng. Methodol.
  \textbf{7}(2),  158--191 (Apr 1998). \doi{10.1145/279310.279314},
  \url{http://doi.acm.org/10.1145/279310.279314}

\bibitem{scikit-learn}
Pedregosa, F., Varoquaux, G., Gramfort, A., Michel, V., Thirion, B., Grisel,
  O., Blondel, M., Prettenhofer, P., Weiss, R., Dubourg, V., Vanderplas, J.,
  Passos, A., Cournapeau, D., Brucher, M., Perrot, M., Duchesnay, E.:
  Scikit-learn: Machine learning in {P}ython. Journal of Machine Learning
  Research  \textbf{12},  2825--2830 (2011)

\bibitem{pennington2014glove}
Pennington, J., Socher, R., Manning, C.: Glove: Global vectors for word
  representation. In: Proceedings of the 2014 conference on empirical methods
  in natural language processing (EMNLP). pp. 1532--1543 (2014)

\bibitem{platt1999proba}
Platt, J.C.: Probabilistic outputs for support vector machines and comparisons
  to regularized likelihood methods. In: Advances In Large Margin Classifiers.
  pp. 61--74. MIT Press (1999)

\bibitem{2018arXiv180210135R}
{Ronen}, R., {Radu}, M., {Feuerstein}, C., {Yom-Tov}, E., {Ahmadi}, M.:
  {Microsoft Malware Classification Challenge}. arXiv e-prints arXiv:1802.10135
  (Feb 2018)

\bibitem{sculley2010kmeans}
Sculley, D.: Web-scale k-means clustering. In: Proceedings of the 19th
  International Conference on World Wide Web. pp. 1177--1178. WWW '10, ACM, New
  York, NY, USA (2010). \doi{10.1145/1772690.1772862},
  \url{http://doi.acm.org/10.1145/1772690.1772862}

\bibitem{searles2017parallelization}
Searles, R., Xu, L., Killian, W., Vanderbruggen, T., Forren, T., Howe, J.,
  Pearson, Z., Shannon, C., Simmons, J., Cavazos, J.: Parallelization of
  machine learning applied to call graphs of binaries for malware detection.
  In: 25th Euromicro International Conference on Parallel, Distributed and
  Network-based Processing (PDP). pp. 69--77. IEEE (2017)

\bibitem{shervashidze2011weisfeiler}
Shervashidze, N., Schweitzer, P., Leeuwen, E.J.v., Mehlhorn, K., Borgwardt,
  K.M.: Weisfeiler-lehman graph kernels. Journal of Machine Learning Research
  \textbf{12}(Sep),  2539--2561 (2011)

\bibitem{siglidis2018grakel}
Siglidis, G., Nikolentzos, G., Limnios, S., Giatsidis, C., Skianis, K.,
  Vazirgiannis, M.: Grakel: A graph kernel library in python. arXiv preprint
  arXiv:1806.02193  (2018)

\bibitem{sutskever2014sequence}
Sutskever, I., Vinyals, O., Le, Q.V.: Sequence to sequence learning with neural
  networks. In: Advances in neural information processing systems. pp.
  3104--3112 (2014)

\end{thebibliography}

\end{document}